\documentstyle[aps,prb,epsfig]{revtex}
\begin{document}

\title{Twin model for orthorhombic perovskites}

\author{Bas B. van Aken, Auke Meetsma, and Thomas T. M.
Palstra$^*$\nocite{pal}}

\address{Solid State Chemistry Laboratory, Materials Science Center,
University of\\ Groningen, Nijenborgh 4, 9747 AG Groningen, the
Netherlands}
\date{\today}

\twocolumn[\hsize\textwidth\columnwidth\hsize\csname@twocolumnfalse\endcsname

\maketitle
 
\begin{abstract}
We present a detailed single crystal x-ray diffraction study of
twinned orthorhombic perovskites. The diffraction pattern can be
indexed on a $2a_p$x$2a_p$x$2a_p$ lattice, but does not obey cubic
symmetry relations. The data can be modelled in space group $Pnma$
with a twin based on a distribution of the $b$ axis over three
perpendicular cubic axes. This model allows full structure
determination in the presence of up to six twin fractions from
single crystal x-ray diffraction data.

\end{abstract}
]
\section{Introduction}

Transition metal oxides with the perovskite structure exhibit a
large variety of interesting physical phenomena, including high
$T_c$ superconductivity, ferroelectricity, colossal
magnetoresistance, and a variety of spin, charge, and orbital
orderings. Doping can change the magnetic and electronic
properties, which can be reflected in the details of the crystal
structure. We are particularly interested in the structural
response of phenomena of electronic origin, like Jahn-Teller
distortions or orbital order.

Since the metal-oxygen bond lengths and angles determine the
exchange interactions, a thorough knowledge of the structure is
necessary to understand the physics behind these phenomena.
Typically crystal structure information is generated by neutron
powder diffraction (NPD). NPD gives direct information about the
distances of lattice planes and is very accurate in determining
the changes of the lattice parameters with temperature and
pressure. Using a full pattern Rietveld Analysis the atomic
positions can be determined with an accuracy of
$10^{-4}$-$10^{-3}$.

The alternative is single crystal X-ray diffraction (SXD). SXD is
a very powerful experimental application to study the crystal
structure of a large variety of materials. In contrast with powder
diffraction, which maps all crystal planes to one dimension
($d$~spacing, or angle $2\theta$), SXD observes all crystal plane
reflections separately, even if they are related by symmetry. The
relative co-ordinates of the atoms in the unit cell can be
determined from the intensity distribution with the same accuracy
as NPD. The major difference is that NPD is sensitive to the
nucleus, while SXD measures the electron density, which is of
interest to understand the electronic properties.

However, for many perovskites SXD is not used because of
unconventional twinning of the crystals. Twinning can complicate
the structure determination by creating extra reflection spots
and/or superimposing reflections on top of each other. Twinning of
the perovskites is complex as the cubic parent structure allows
not only simple twin axis and planes, but also 3D twin relations.
In this paper we will show a method to analyse twinned single
crystals. Moreover, this twin model allows a full structure
determination including detailed information on distortions of
both structural and electronic origin. Besides, we also show that
one does not need a larger data set, as is usual in the case of
twinning.

Using SXD, we observe a $2a_p$x$2a_p$x$2a_p$ unit cell. It is
commonly accepted that the unit cell is
$\sqrt{2}a_p$x$2a_p$x$\sqrt{2}a_p$. The observed unit cell
originates from a three-dimensional type of twinning that is not
restricted to manganites. The model is most likely of general
application for a large variety of perovskite $Pnma$ crystals.
This twinning is unique and it involves a distribution of the
$b$~axis over three perpendicular cubic axes.

\section{The twin structure for manganites}

The manganites have generated considerable attention because of
the colossal magnetoresistance effect. The role of magnetic order
has been widely discussed as the double exchange, inducing the
ferromagnetic order, is required to generate a metallic ground
state. The role of orbital order is much less understood. While
{\it local} Jahn-Teller distortions are crucial in explaining the
localization of the charge carriers in the paramagnetic state, the
{\it long range} Jahn-Teller ordering is not well studied, except
for undoped LaMnO$_3$. We will show that the destruction of long
range orbital order is a second prerequisite for the metallic
ground state. In fact, we will show in a separate paper that with
increasing hole doping of LaMnO$_3$ not the exceeding of a
critical concentration but the suppression of the orbital order
generates the metallic ground state. The orbital ordering in
perovskites with degenerate $e_g$ electrons can be easily
measured, whereas for degenerate $t_{2g}$ electrons the
Jahn-Teller distortions are much smaller. Furthermore, the $Pnma$
symmetry can accommodate not only the Jahn-Teller ordering, but
also a 3D rotation of the octahedra, known as the GdFeO$_3$
distortion. In this paper we will focus on the main reason why SXD
has not been widely used for these perovskites, namely twinning.
Twinning in doped LaMnO$_3$ originates from the transition of the
highly symmetric cubic parent structure to the orthorhombic
symmetry, that accommodates both the GdFeO$_3$ and the Jahn-Teller
distortion. In thin films $ac$ twinning has been observed. The
solution of the twin relations in the crystals allow us to study
in detail the ordering of these compounds as influenced by
temperature, magnetic state and doping concentration. In this
paper we will focus on the twin relations.

\medskip
At high temperatures most AMnO$_3$ are cubic with a unit cell of
$a=3.9$ \AA. Due to both the small average A-site radius and the
JT effect the MnO$_6$ octahedra are rotated and distorted at lower
temperatures. We will show that our crystals are twinned. We
present here an accurate description of the origin of the
twinning, the detwinning process and show that we can still
determine the temperature dependence of the structure by measuring
only the reflections of the main fraction in one octant of the
$hkl$ space.

Twinning is reported before in AMnO$_3$ perovskites and related
structures. For instance, neutron powder experiments on LaMnO$_3$
showed reflections, at temperatures above the Jahn-Teller
transition, that could be indexed on a double cubic unit cell. But
full pattern refinement was only possible in the orthorhombic
space group $Pnma$.\cite{Rod98} Single-crystal electron
diffraction showed a double cubic unit cell for SrSnO$_3$, even
though CaSnO$_3$ is orthorhombic.\cite{Veg86} Electron microscopy,
however, showed the existence of coherent domains, with
perpendicular orientations of the doubled $b$ axis. The structure
was deduced via the O'Keefe-Hyde relations.\cite{Oke77}

The single crystal was mounted on an ENRAF-NONIUS CAD4 single
crystal diffractometer. The temperature of the crystal was
controlled by, heating, a constant nitrogen flow. Initial
measurements were done at 180 K. Temperature dependent
measurements were performed on a diffractometer with an adjustable
temperature set-up, between 130 K and 300 K.

\section{Crystal structure}
\fnsymbol{footnote}

Most structure research of perovskites focuses on the Mn-O
distances and the Mn-O-Mn angles, as these parameters determine
the super- and double exchange interactions. Here, we like to
stress the importance of a complete structure refinement,
including the La-position and the rotation/distortion of the
MnO$_6$ octahedra. The basic deformations that determine the
deviation from the cubic structure are the GdFeO$_3$ distortion
and the Jahn-Teller distortion.

The basic building block of the perovskite structure is a 3.9
{\AA} cube with Mn in the centre and O at the face-centers. The
oxygen ions co-ordinate the Mn to form MnO$_6$ octahedra. The A
atoms are located at the corners of the cube. The undistorted
'parent' cubic structure rarely exists, but distorts
conventionally to an orthorhombic or rhombohedral symmetry. For an
ideal perovskite the ratio between the radii of the A site ion and
the transition metal ion is such that the tolerance factor
\begin{equation}
t=\frac{\langle r_{A^{2/3+}}\rangle+r_{O^{2-}}}{\sqrt2(\langle
r_{Mn^{4/3+}}\rangle+r_{O^{2-}})}
\end{equation}
is equal to one. In the La-Ca system the tolerance factor varies
from 0.943 to 0.903 going from LaMnO$_3$ to CaMnO$_3$.

\subsection{GdFeO$_3$ rotation} Due to the small radius of the A-site ion, with respect
to its surrounding cage, the MnO$_6$ octahedra tilt and buckle to
accommodate the lanthanide. This is known as the GdFeO$_3$
distortion. The cubic state allows one unique oxygen position. Due
to the GdFeO$_3$ distortion we need two inequivalent positions to
describe the structure. O1 is the in-plane oxygen, on a general
position, $x,y,z$. Two opposite Mn-O1 bonds have the same length,
but the perpendicular bonds need not to be equal. O2 is the apical
oxygen, located on a fourfold $x,\frac{1}4,z$ position on the
mirror plane. Mn-O2 bonds are always of the same length. Both in
the undistorted and the distorted perovskite, the O-Mn-O bond
angles are $180^\circ (90^\circ)$, but due to the buckling Mn-O-Mn
bond angles are no longer $180^\circ$. A pure GdFeO$_3$ distortion
can be obtained with equal Mn-O bond lengths.

\subsection{Jahn-Teller distortion}

The Jahn-Teller effect originates from the degenerate Mn $d^4$
ion. Two, possible, distortions are associated with the
Jahn-Teller effect. Q2 is a orthorhombic distortion, with the
in-plane bonds differentiating in a long and a short one. Q3 is
the tetragonal distortion with the in-plane bond lengths
shortening and the out-of-plane bonds extending, or vice
versa\cite{Kan60,Yam96}. The main result of the JT effect, Q2 and
Q3, is that the Mn-O distances become different, which lifts the
degeneracy of the $t_{2g}$ and $e_g$ levels.

\subsection{Glazer's view on the octahedra}
There are three possible rotations for a rigid MnO$_6$ octahedron.
One, $\eta_y$, changes the atomic positions of the O2 atom only.
The others have an effect on the atomic positions of both O1 and
O2. These movements will also affect the position of the La
atom\cite{Miz99}. Glazer\cite{Gla72} identified all possible
sequences of rotations for the perovskite system. Two of these
sequences create the symmetry elements that make up the standard
perovskites. In Glazer's notation: $a^-b^+a^-$ \footnote{$^)$
$p^-q^+r^0$ means that along [100] a rotation of size $p$ is
alternated positive and negative, along [010] a rotation of size
$q$ is always (arbitrarily) positive, and along [001] there is no
rotation.}$^)$ yields orthorhombic $Pnma$ and $a^-a^+a^-$ will
result in rhombohedral $R\overline{3}c$. As Glazer worked in a
pseudocubic $2a_p$x$2a_p$x$2a_p$ system we have to translate these
rotations to the orthorhombic unit cell, which gives $a^-b^+0$.
This means that the rotation around the $c$~axis, $\eta_z$, is
zero. This is in agreement with the observation that the four
positions of O2 near the Mn at 0,0,0 have the same distance to the
$y=0$ plane, albeit two are positive and two are negative. By
rotating the MnO$_6$ octahedra around the $x$~axis, two oxygen are
raised from their cubic position and two are lowered. Here we
neglect the influence of a rotation $\eta_y$ on these coordinates.
As a result of $\eta_z=0$ the fractional $x$~co-ordinate of O1 is
also zero. In practice small deviations are found, indicative of
the non-rigid behaviour of the MnO$_6$ octahedra. Typically in
AMnO$_3$ two rotations and the Q2 JT distortion are observed.

One effect of the GdFeO$_3$ distortion is that the number of
formula units per unit cell is enlarged. The unit cell is doubled
in the $b$~direction, with respect to the original cubic cell. The
$ac$~plane is also doubled, resulting in $b\approx2a_p$, and
$a\approx c\approx \sqrt{2}a_p$. Despite the rotations and
distortion, the structure remains in origin cubic. The shifts of
the atoms in the unit cell are small. Furthermore, Mn stays
roughly octahedrally surrounded although the Mn-O distances tend
to differ and some O-Mn-O angles may be no longer perfectly
$90^{\circ}$ or $180^{\circ}$. The small deviations from the cubic
symmetry are reflected in the intensities of the reflections. To
illustrate this point, we transform an arbitrary, cubic,
reflection $hkl$ to the orthorhombic setting by $h'k'l'$ =
$h$+$l$,$2k$,$h$-$l$. Thus orthorhombic reflections with $k'$ even
and $h'$ and $l'$ both even or both odd stem from planes that
already existed in the original cubic unit cell. We observe in
diffraction patterns of orthorhombic perovskites that reflections
that originate from cubic crystal planes have higher intensities
than those that do not.

\section{Twin model}

\begin{figure}[htb]
\centering
\includegraphics[width=80mm,height=68.2mm]{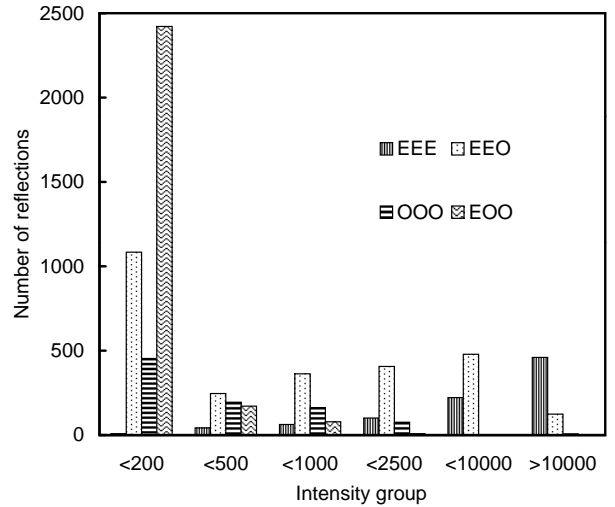}
\caption{The number of reflections of the different types versus
intensity ranges. Low intensities, $I<200$ are mostly
\textsc{eoo}, while high intensities, $I>2500$, are of the types
\textsc{eee} and \textsc{eeo}.} \label{all}
\end{figure}

Reflections in reciprocal space are experimentally observed at a
regular distance in three orthogonal directions, corresponding to
a cubic lattice spacing of $7.8$ \AA\ in real space. Although the
three orthonormal axes have equal lengths, we could not observe
the threefold rotation axis along $\langle111\rangle$, required
for cubic symmetry. However, studying the intensity distribution
of planes in $hkl$ space with constant $h$, $k$ or $l$ showed much
regularity as is shown in Fig.~\ref{all}. We propose a twinned
structure consisting of coherent $Pnma$ domains. Due to partial
overlap this results in a metric cubic system with
$a\approx2a_p\approx7.8$~\AA.

\medskip
Twinning is often observed in crystals with a reduced symmetry.
For instance, the orthorhombic perovskite LaMnO$_3$ has lower
symmetry than cubic SrTiO$_3$. Conventional twin models keep one
characteristic axis unchanged and form the domains by rotation
around that axis. This is commonly observed in constrained
epitaxial thin films and pseudo-two-dimensional crystals like
YBa$_2$Cu$_3$O$_{7{-}\delta}$. Another standard twin is inversion
twinning, found in non-centrosymmetric systems. Due to the
inversion twin, pseudo-centrosymmetry is obtained or a net
polarisation in ferroelectric compounds is reduced to
zero.\cite{Rao97a,Van01a} Both twin models have usually the
property that either all twin domains have reflections that lie on
top of each other, merohedral twins, or there is a
non-commensurate set of reflections, {\it e.g.} in monoclinic unit
cells. Here we propose a more extensive form of twinning, where
the different reflections of different domains partially coincide,
giving rise to the observed metrically cubic system.

\begin{figure}[tb]
\centering
\includegraphics[width=70mm,height=140mm]{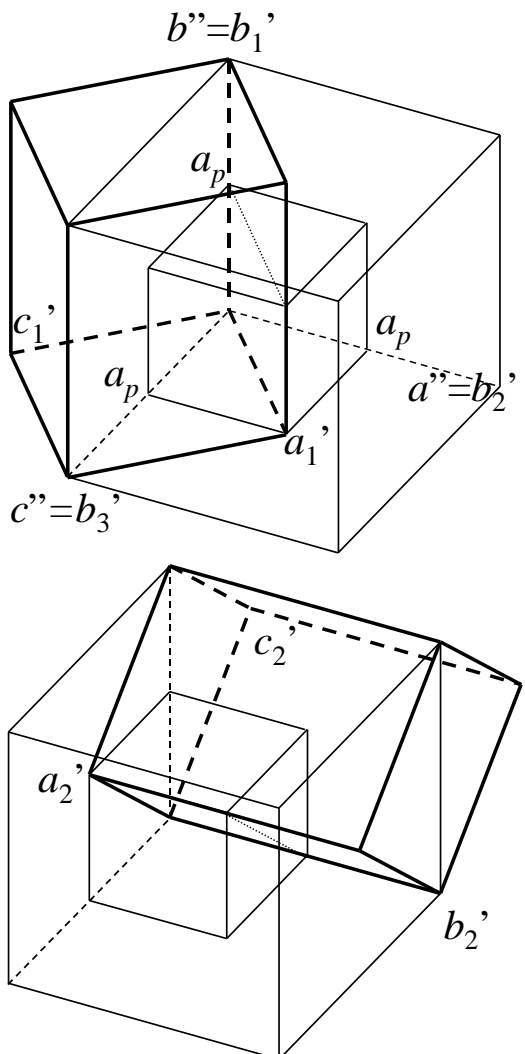}
\caption{Two of three possible orientations of the $Pnma$ unit
cell in the observed $2a_p$x$2a_p$x$2a_p$ ($a''$x$b''$x$c''$) unit
cell. $b_1'$, $b_2'$ and $b_3'$ denote the orientations of the
doubled axes along the original cubic axes, $a_p$. Note that the
choice of $a'$ and $c'$ is still open, after choosing an
orientation for the $b'$ axis.} \label{2of3}
\end{figure}

\medskip
The transformation of cubic to orthorhombic symmetry requires a
designation of $a$, $b$, and $c$ with respect to the degenerate
cubic axes. There are three possibilities to position the doubled
$b$~axis along the three original cubic axes. Thus we propose that
the three fractions' $b$~axes are oriented along the three
original axes of the cubic unit cell, as sketched in
Fig.~\ref{2of3}. This twin model consists of a $Pnma$ unit cell,
transformed by rotation along the 'cubic' [111] axis. We still
have the freedom to choose the $a$ and $c$~axes, perpendicular to
the $b$~axis, and rotated 45$^\circ$ with respect to the cubic
axes. Therefore, this model yields six different orientations of
the orthorhombic unit cell. As the differences between
$a/\sqrt{2}$, $b/2$ and $c/\sqrt{2}$ are small, we observe the
reciprocal superposition of the six orientations as metrically
cubic.

The standard refining programs can work with twin models, but only
if they consists of merohedral twins. In our case, the reflections
do not correspond with an orthorhombic unit cell, but with a,
twice as large, cubic unit cell. In appendices
\ref{app:extinctions} and \ref{app:model}, we will elaborately
explain the transformation from orthorhombic to double cubic and
the way we worked around the refining program.

We first measured a hemisphere in $hkl$ space with -$20<h<20$,
-$20<k<20$, and $0<l<20$. On this dataset, we could refine our
twin model with the six fractions, including the Ca concentration
on the A~site. Measuring a large range in $hkl$ space requires a
large amount of time, roughly 10 days in the present set-up.
Conventionaly, for crystals with orthorhombic unit cells measuring
one octant is sufficient. From refinements on selected parts of
the dataset, we concluded that we could investigate the structure
at different temperatures by only measuring the positive octant of
the main fraction, thereby limiting the measuring time
considerably.

\section{Discussion}
\begin{figure}[tb]
\centering
\includegraphics[width=80mm,height=62.7mm]{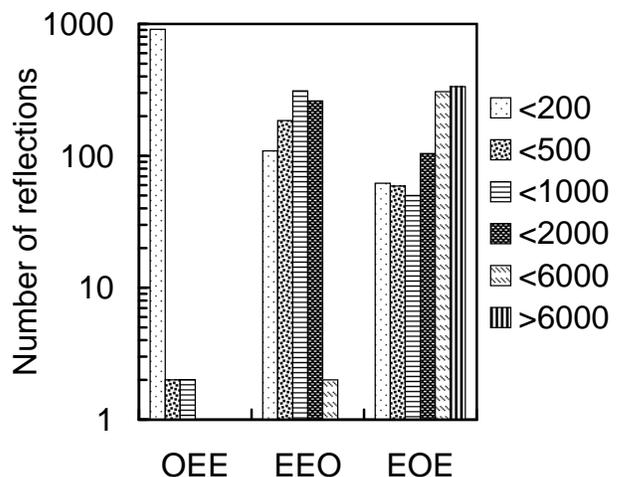}
\caption{The number of reflections versus the types \textsc{oee},
\textsc{eeo} and \textsc{eoe} for different intensity ranges.
\textsc{oee}, $b'$ parallel to $a''$, has mostly very low
intensity. Reflections \textsc{eoe}, $b'$ parallel to $b''$, have
a broad range of intensities. In cubic symmetry, these
distributions should be identical for absorption corrected data,
which we considered.} \label{eeo}
\end{figure}

The observed reflections could be indexed on a cubic lattice with
$a\approx7.8$ \AA. Although the three axes had equal lengths, we
could not observe the required threefold symmetry axis along
$\langle111\rangle$ for cubic symmetry. This is shown for
\textsc{eeo} reflections in Fig.~\ref{eeo}. Furthermore, studying
the intensity distribution of $hkl$ planes with constant $h$, $k$
or $l$ showed much regularity, as is shown in appendix
\ref{app:extinctions}. Rodr$\mathrm{\acute{i}}$gruez-Carvajal {\it
et al.} noted that their neutron powder spectra of pure LaMnO$_3$
could be indexed in a cubic $2a_p$x$2a_p$x$2a_p$ unit cell, above
the Jahn-Teller transition temperature.\cite{Rod98} They did not
observe a splitting of the peaks. Nevertheless, they could not
refine the structure in a cubic unit cell. Note that in powder
diffraction spectra there is no direct method to observe the
threefold symmetry along the body diagonal. The refinement of the
spectra has been done in the conventional $Pnma$ setting. They
suggested that the powder is twinned but could not give evidence
for that, as they only studied powders. The observed reflections,
the systematic extinctions and the non-cubic intensity
distribution suggest that our single crystals are also twinned.

Typically, Ca-doped LaMnO$_3$ has space group $Pnma$. We propose a
twinned structure consisting of coherent $Pnma$ domains, which
results in a metric cubic system with $a\approx7.8$~\AA. The model
is identical to the twinning observed by electron microscopy in
CaSnO$_3$.\cite{Veg86} They could however not refine their
$2a_p$x$2a_p$x$2a_p$ data on different $Pnma$ domains, and deduced
the structure from phenomenological relations\cite{Oke77} and the
similarities with SrSnO$_3$. In the present work, the different
contributions of the twin fractions to the total integrated peak
intensity are taken into account. The refined model consisted of
the regular parameters in single crystal diffractometer. We
considered for each observed reflection the related orthorhombic
reflections for the applicable twin fractions and used an
identical unit cell. This way, we can refine both the atomic
positions of the asymmetric unit and the volume ratio of the twin
domains simultaneously.

\medskip
Several crystals were measured and refined with this detwin model.
The refinements showed that for every crystal the distribution of
the volume over the twin fractions is different. This is an
indication that our crystals are really twinned. Other structural
deformations would lead to a constant, not sample dependent,
effect on the structure factors and therefore to the same
fractions of the twin domains. We can also conclude that the size
of the twin domains must be slightly smaller than the magnitude of
the measured crystals, i.e. several tenths of a millimeter. Larger
twin domains would give rise to crystals of one single domain and
smaller domains are more likely to produce a constant spreading of
the domains. Non-regular, though constant, distributions could
signal a preferential growth direction of the crystal.

\subsection{Intensity distribution}
Careful analysis of the intensity distribution showed extra
evidence for the proposed models. The full dataset shows that
reflections $hkl$ of type \textsc{eee}\footnote{$^)$\textsc{e}
denotes an even value for the Miller indices, \textsc{o} stands
for odd}$^)$ are by far the strongest. Reflections having one odd
Miller index, \textsc{oee}, \textsc{eoe}, \textsc{eeo}, are second
in intensity. \textsc{eoe} has the highest intensity of these
reflections as seen in Fig.~\ref{eeo}. To understand the intensity
distribution in the $2a_p$x$2a_p$x$2a_p$ unit cell, we examine the
structural deformations. Despite the effect of Jahn-Teller
distortions, GdFeO$_3$ rotations and twinning, the structure
remains in origin cubic. This means that the intensity
distribution of the peaks with high intensity will always mimic
the intensity distribution of the peaks of the undistorted cubic
perovskite. An arbitrary reflection, in the cubic unit cell,
$hkl$, is transformed to the orthorhombic unit cell as $h+l$,
$2k$, $h-l$. Thus orthorhombic reflections with $h'$ and $l'$ both
even or both odd and $k'$ even stem from the original cubic unit
cell. Although the fact that the $h'k'l'=$\textsc{eoe} reflections
do not originate from the original cubic planes, they are allowed
in the $Pnma$ symmetry. Reflections $h'k'l'$ that satisfy these
conditions have higher intensities than those that do not.

To double the $Pnma$ unit cell to a $2a_p$x$2a_p$x$2a_p$ unit
cell, a reflection $h'k'l'$ is transformed to $h''k''l''$ as
$h'+l'$, $k'$, $h'-l'$. Please note that we do not explicitly
mention the different orientations of the orthorhombic unit cell.
If the regarded $b$~axis is not parallel with $k''$, than the
indices should be cycled appropriately. The transformation yields
that double cubic reflections with $k''=k'=2n$ originate from
single cubic reflections, with $k=n$. However, $k''=2n+1$
reflections originate from single cubic non-integer indices $k$,
and will therefore have less intensity. This explains why the
measured $h''k''l''=$\textsc{eee} reflections are strongest and
\textsc{eoe} are one order of magnitude less strong. We have seen
that $h''=h'+l'$ and $l''=h'-l'$ thus $h''$ and $l''$ are always
both even or both odd. Therefore \textsc{eoe} has only
contribution of the fractions with $k''$ corresponding to the
orthorhombic $b$~axis. \textsc{oee} and \textsc{eeo} originate
from fractions with their $b$~axis parallel to $h''$ and $l''$,
respectively. The fact that our observed \textsc{eoe} is stronger
than \textsc{oee} and \textsc{eeo}, indicates that, by chance, the
observed $b$~axis, or $k''$~direction, is parallel with the
$b$~axis or $k'$~direction of the largest twin fraction.

\subsection{Refinement}
We determined both the Ca content and the volume fractions of the
twins by measuring a full hemisphere of $hkl$ space. We compared
the results with a refinement on only one octant of these measured
reflections. The outcome was equivalent within the error bars. We
refined again fixing the twin fractions and Ca concentration to
the values found for the first refinement on the large dataset,
and allowing only the atomic co-ordinates and anisotropic
displacement parameters (adp's) to change. This resulted in the
same structural model as found with the complete structure
determination on the largest dataset. We concluded that apparently
the refinement with fixed fractions and Ca concentration is
insensitive for the size of the dataset. This can be understood if
we view the symmetry relations and Friedel pairs. Due to the
Patterson symmetry, $mmm$ the relation $F(\overline{h}kl)=F(hkl)$
is valid for all structure factors. Friedel pairs are reflections
that have intrinsically the same structure factor, related thus by
symmetry and ignoring absorption and anomalous scattering effects.
Although we have six different orientations, one octant chosen for
a particular orientation will still contain data of all fractions
that can be transformed in such a way that every $h'k'l'$ is
included for all three $b$~directions.

\section{Conclusion}
We have shown that crystals of La$_{1-x}$Ca$_x$MnO$_3$, with
$x=0.19$, have the orthorhombic $Pnma$ space group, although they
appear in single crystal diffractometry metrically cubic. The
metric cubic appearance results from twinning in which the doubled
$b$~axis of the $Pnma$ cell is oriented along all three cubic axes
of the single cubic unit cell during crystal growth or cooling.
This twin model allows successful structure refinements of single
crystal x-ray diffraction data.

\appendix
\section{Data analysis}
\label{app:extinctions} Here we describe the effect of the twin
model on the visibility of the standard $Pnma$ reflection
conditions as well as a detailed analysis of the intensity
distribution of a model crystal with $Pnma$ symmetry and of the
measured reflections in double cubic setting. The first step in
space group determination is to look for systematic absences in
the list of reflections. We observe no intensity for the
reflections shown in table \ref{t1}. Note that the indices are in
double cubic setting and therefore allow cyclic permutation, {\it
i.e.} $h00$, $0k0$ and $00l$ are all represented by $h00$. These
reflection conditions allow only the space groups $P2_13$ and
$P4_232$. However, these space groups have as reflection condition
only $h00$ : $h=2n$. This suggests that we have more information
than can be attributed to these space groups. Attempts to
determine the structure using these space groups were
unsuccessful. Furthermore, the observed and, for cubic systems,
anomalous, reflection conditions suggested that we might be
looking at a twinned crystal. In the next paragraph, we present
the transformation of the reflection conditions for orthorhombic
$Pnma$ according to the presented twin model.

\begin{table}[htb]
\centering \caption{Extinct reflections as observed in the
$2a_p$x$2a_p$x$2a_p$ data set.}
\begin{tabular}{c l l c}
 & Reflection&Extinction condition &\\
 &$h00$           & $h\neq2n$       & \\
 &$h\overline{h}0$& $h\neq2n$       & \\
 &$hkh$           & $h\neq2n, k=2n$ & \\
 &$hk\overline{h}$& $h\neq2n, k=2n$ & \\
\end{tabular}
\label{t1}
\end{table}

\subsection{Extinctions}
The $Pnma$ reflection condition $0k'0$ : $k'=2n$ is transformed to
double cubic as $0k''0$ : $k''=2n$ and cyclic permutation. This
corresponds to the reflection condition $h''00$ : $h''=2n$ in
table \ref{t1}. $Pnma$ reflections of the type $h'0h'$ are
transformed to $h''00$, with $h''=2h'$. This implies that $h'0h'$
never contributes intensity on reflections $h''00$, with $h''$
odd. Conversely, if we do not observe intensity on all reflections
$h''00$ : $h''\neq2n$, then all of the constituting $Pnma$
reflections should be absent.

\smallskip
To transform and understand the other reflection conditions of
$Pnma$ we have to take into account that an orthorhombic, and also
a cubic, unit cell implies that $hkl$ is equivalent upon sign
reversal of each of the Miller indices. The $Pnma$ reflection
condition $h'00$ : $h'=2n$ becomes $h''0h''$ : $h''=2n$ and
similarly $00l'$ transforms to $l''0\overline{l''}$. But as we
just stated, $h''0h''$ is identical to $h''0\overline{h''}$. If we
combine reflections then a reflection condition is fulfilled if
any of the contributing parts has intensity. To obtain extinction,
all contributing reflections should be extinct. As $00l'$ has no
extinction condition, {\it i.e.} intensity for all $l'$, they will
contribute to all $h''0h''$/$h''0\overline{h''}$ reflections, and
the $Pnma$ reflection condition $h'00$ : $h'=2n$ is masked.

\smallskip
The overlap of different reflections occurs also for $0k'h'$ and
$h'k'0$. These transform to $h''k''\overline{h''}$, and
$h''k''h''$. Therefore, we cannot disentangle $0k'h'$ and $h'k'0$
in the measured double cubic cell. The $Pnma$ reflection
conditions for $0k'l'$ and $h'k'0$ are $k'$+$l'=2n$ and $h'=2n$,
respectively. $h''k''h''$ and $h''k''\overline{h''}$ are found to
be extinct for $h''\neq2n$ and $k''=2n$ in the double cubic
setting. To have these reflection extinct, all contributing $Pnma$
reflection conditions must be extinct. This implies that
\emph{both} $h'=2n$ \emph{and} $k'$+$h'=2n$, the reflection
conditions for the contributing $0k'h'$ and $h'k'0$, may not be
fulfilled. Thus $h''\neq2n$ and this yields $k''=2n$.

\medskip Now we consider reflections of the type $h''k''h''$ that
are not extinct. Our twin model yields six fractions that can
contribute to the considered reflections.

\begin{enumerate}
\item
$h''=2n$ and $k''=2n$. Reflections \textsc{eee} have contributions
from the four fractions with the doubled $b'$-axis parallel to
either $a''$ or $c''$, this is also true for reflections
\textsc{ooo}. In addition, we also have contributions from the two
fractions with $b'$ parallel to $b''$, \emph{i.e.} from $h'k'0$
and $0k'h'$.
\item
$h''\neq2n$ and $k''=2n$. Reflections \textsc{ooo}, in addition to
the four fractions with the doubled $b'$-axis parallel to either
$a''$ or $c''$, have one additional contribution from $0k'h'$. In
the special case that $h''=k''$, for both \textsc{eee} and
\textsc{ooo}, then two of the four fractions, parallel to $a$ and
$c$ contribute.
\item
$h''=2n$ and $k''\neq2n$. Reflection \textsc{eoe}, have only a
contribution from the $h'k'0$ reflection, with $b'$ parallel to
the observed $b''$~axis.
\end{enumerate}

We have shown that the intensity of observed $h''k''h''$
reflections depend on odd or even. We conclude that $h''k''h''$
reflections are extinct for $OEO$, while the other reflections
have contributions from up to six twin fractions.

\subsection{Intensity in double cubic setting compared with $Pnma$}

We can also learn something about the twinned origin by studying
the intensity distribution. Our twin model provides a natural
explanation for the hierarchy in intensities of the reflections.
First, we consider the known transition from the single cubic unit
cell to the orthorhombic one to elucidate the patterns that are
inherent to perovskites. Then we proceed to show how these
patterns are convoluted in the double cubic, twinned model.

\begin{figure}[t]
\centering
\includegraphics[width=80mm,height=45mm]{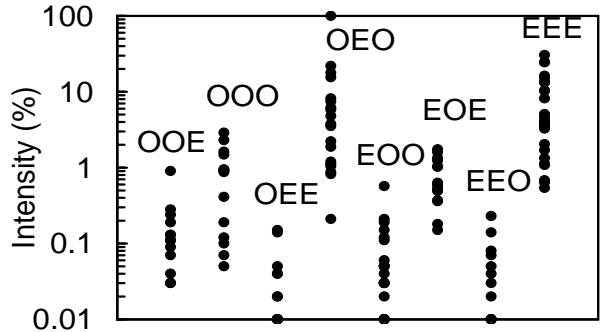}
\caption{The calculated intensities for a conventional perovskite,
subdivided with respect to odd and even of the Miller indices.
$k=2n$ reflections are strongest, if $h'+l'=2n$. Reflections with
$h'+l'\neq2n$ have roughly 100 times less intensity.} 
\label{Pnma}
\end{figure}

\smallskip
The intensity distribution of a normal $Pnma$ perovskite is shown
in Fig.~\ref{Pnma}. In conventional orthorhombic $Pnma$ the
reflections $h'k'l'$ with $h'+l'=2n$ are stronger than
$h'+l'\neq2n$. From these, the $k=2n$ reflections are stronger
than the $k\neq2n$ ones. This can be attributed to their origin in
the 3.9 \AA\ cubic structure. The $h'k'l'$ reflections in
orthorhombic setting originate from $hkl$ in the single cubic cell
for $h'=h$+$l$ and $l'=l$-$h$. Therefore, $h'$ and $l'$ will
always be both even or both odd. In other words if $h'$-$l'\neq2n$
than the reflections originate from a crystal plane in the single
cubic setting with a non-integer Miller index, \emph{i.e.} a
superlattice reflection and their intensities are usually weak.
The same argument can be applied to $k$. As $k'=2k$, thus $k'=2n$
originate from a regular crystal plane, $k'\neq2n$ is the
superlattice reflection. These $k'\neq2n$ however appear to have
somewhat more intensity than the $h'$-$l'\neq2n$ reflections.
\smallskip
\bigskip
\smallskip

We observe the following patterns in the intensity distribution in
the double cubic cell. First, reflections \textsc{eee} are
generally the strongest, \textsc{eoo} usually absent, or very
weak. If we ignore the allowed permutations in cubic setting, the
following hierarchy can be made. \textsc{eoe} reflections were
much stronger than \textsc{eeo}, which are again stronger than
\textsc{oee}. Likewise, \textsc{oeo} reflections were, although
weak, stronger than \textsc{ooe} and again stronger than
\textsc{eoo}.

In conventional $Pnma$ structure we found that if $h'$+$l'\neq2n$
than reflections with $k'\neq2n$ are stronger than those with
$k'=2n$. In our notation, for the $Pnma$ structure this yields
(\textsc{eee}=\textsc{oeo}) $>>$ (\textsc{ooo}=\textsc{eoe}) $>>$
(\textsc{ooe}=\textsc{eoo}) $>$ (\textsc{oee}=\textsc{eeo}). So
there are roughly four intensity groups. We can correlate these
four groups with the four types of reflections we measured in the
supercubic setting, with the following intensity hierarchy:
\textsc{eee}$>>$\textsc{eeo}$>>$\textsc{ooo}$>$\textsc{eoo}.

We consider here the transformation from the orthorhombic to the
double cubic setting for the in-plane indices, $h'$ and $l'$.
Double cubic indices are calculated with $h'$+$l'$ and $h'$-$l'$,
therefore $h''$ and $l''$ will be both even or both odd.
Reflections of the type \textsc{e?o} or \textsc{o?e} will have no
intensity of the twin with $b'$ parallel to $b''$, as they are not
originating from an 'integer' crystal plane. But we have a mixing
of different orientations of the $b'$~axis along the three cubic
axes $a''$, $b''$ and $c''$. We can cycle the indices of these
reflections such that we get either \textsc{e?e} or \textsc{o?o},
and from the corresponding twin, $b'$ parallel to $a''$ or $c''$
we do have intensity on these reflections. \textsc{eoe} and
\textsc{oeo} will only have contributions from this particular
setting. \textsc{eee} and \textsc{ooo} can be cycled and in
general have contributions from all six orientations.

\subsection{distribution of the twins} If we consider that the
main part of the intensity in the double cubic setting originates
from the largest twin fraction, we can search for the orientation
of the largest fraction. That $b'$~axis is expected to be oriented
along one of the measured double cubic axis. We can not
differentiate the two fractions with parallel $b'$ and
perpendicular $a'$ and $c'$. We considered groups of reflections
with cycled indices, having either one odd, \textsc{oee}, or one
even index, \textsc{eoo}. Roughly, the intensities of the
reflections within these groups occurred with ratio 70:20:10. This
suggested that one twin had 70\% of the volume, the others 20\%
and 10\%. \smallskip Now we sorted these reflections to find the
corresponding orientations. The sort parameter is the observed
double cubic axis that corresponded with the odd indices in
\textsc{oee}, and vice versa. Only cycled reflections that
occurred three times with measurable intensity were taking into
account. We found that $I_{a''}:I_{b''}:I_{c''}=5:80:15$. The
constraint that all three intensities have positive values,
ignores the weakest reflections. It also ignores Bijvoet pairs, as
we only measured $-20<l''<2$. If we indeed had a cubic system then
these variations should be zero within standard deviation. The
distribution of the intensities with respect to the different
axis, strongly suggests that the crystal is twinned.

\section{Implementation in refinement}
\label{app:model}

Here we describe the model and the application by SHELXL. The
standard refining programs can work with merohedral twin models.
However, in our case, the reflections do not coincide with one
orthorhombic unit cell. Every observed reflection has
contributions of up to six twin fractions. We measured both in the
double cubic setting as in the orthorhombic setting of the main
twin fraction. We transformed the $h''k''l''$ indices to the six
possible twin orientations. Three of the possibilities are given
by
\begin{eqnarray}
h'k'l'=\frac{1}{2}(h''+l''), k'', \frac{1}{2}(h''-l'') \\
h'k'l'=\frac{1}{2}(l''+k''), h'', \frac{1}{2}(l''-k'') \\
h'k'l'=\frac{1}{2}(k''+h''), l'', \frac{1}{2}(k''-h'')
\end{eqnarray}
The other three can be acquired by changing $h'$ for $l'$ and $l'$
for -$h'$.

A new software program TWINSXL was developed to transform the
standard data file, HKLS, by using the appropriate transformation
matrices from a second input file.\cite{Mee00} The new data file
constitutes of lines of $h, k, l$, intensity plus standard
deviation and twin fraction number. We used the "HKLF 5" option of
SHELXL to refine data. The refinement uses a crystal model for the
orthorhombic structure with the normal, adjustable variables. Five
variables for the twin fractions were added, the sixth fraction is
calculated as the complement of the other five fractions. The sum
of the appropriate calculated intensities for all fractions was
compared with the observed integrated intensities.

\end{document}